\documentclass[aps, llncs, preprint, nofootinbib, showkeys]{revtex4-1} 
\usepackage{amsmath}
\usepackage{amsfonts}
\usepackage{amssymb}
\usepackage{graphicx}
\usepackage{float}
\usepackage[export]{adjustbox}
\usepackage{ragged2e}
\usepackage[utf8]{inputenc}
\usepackage[T1]{fontenc}

\makeatletter
\newcommand{\printfnsymbol}[1]{%
 \textsuperscript{\@fnsymbol{#1}}%
}
\makeatother
\begin{document}

\title{Electrically controlled quantum transition to an anomalous metal in 2D}

\def\thefootnote{$^a$}\footnotetext{These authors contributed equally to the work.}

\author{Soumyadip$^{a}$ Halder$^{1}$, Mona$^{a}$ Garg$^{1}$, Shreekant Gawande$^1$, Nikhlesh Singh Mehta$^1$, Anamika Kumari$^2$,  Suvankar Chakraverty$^2$, Sanjeev Kumar$^1$}
\author{Goutam Sheet$^1$}
\email{goutam@iisermohali.ac.in}
\affiliation{$^1$Department of Physical Sciences, Indian Institute of Science Education and Research (IISER) Mohali, Sector 81, S. A. S. Nagar, Manauli, PO 140306, India}

\affiliation{$^2$Quantum Materials and Devices Unit, Institute of Nano Science and Technology,
Sector-81, S. A. S. Nagar, Punjab, 140306, India.}

\begin{abstract}
    The mechanism through which superconductivity is destroyed upon controlled disordering often holds the key to understanding the mechanism of emergence of superconductivity. Here we demonstrate an $in$-$situ$ mechanism to control the fraction of disorder in a 2D superconductor. By controlling an electric field $V_G$, we created an assembly of segregated superconducting nano-islands and varied the inter-island distance to accomplish a quantum phase transition from a superconducting phase to a strange quantum anomalous metallic (QAM) phase at LaVO$_3$/SrTiO$_3$ interfaces. In the QAM phase, the resistivity dropped below a critical temperature ($T_{CM}$) as if the system was approaching superconductivity, and then saturated, indicating the destruction of global phase coherence and the emergence of a phase where metal-like transport of Bosons (a Bose metal) becomes a possibility. The unprecedented control over the island size is obtained through the control of nanometer scale ferroelectric domains formed in the SrTiO$_3$ side of the interface due to a low-temperature structural phase transition. 
    \end{abstract}
\maketitle

\section*{Introduction}
In three dimensions (3D), the vast majority of clean metals with weak interactions are described well by Landau's Fermi liquid theory\cite{baym_landau_1991}. Within this theory, electronic transport is facilitated by the fermionic quasiparticles that undergo inelastic scattering from various scattering centers or disorders in lattice periodicity, thus causing resistance. In a completely different paradigm, under favourable conditions, in presence of an effective attractive interaction between two such fermions, the fermions pair-up to form bosonic Cooper pairs which can all populate the ground state of the system leading to a coherent charge transport that is dissipation-less, superconducting, even in presence of disorder\cite{bardeen_theory_1957}. As per the fundamental understanding, while nonmagnetic impurities should not destroy superconductivity in 3D, an adequately disordered electronic system must be an insulator due to Anderson localization\cite{anderson_absence_1958}. This contradiction makes the disorder-induced transition from a superconducting to an (Anderson) insulating phase an extremely interesting area of research\cite{haviland_onset_1989,paalanen_low-temperature_1992, sacepe_disorder-induced_2008, goldman_superconductor-insulator_2010}. 

In two-dimensions (2D), the science is even more fascinating, where within the Fermi liquid theory itself, a metal should turn into an Anderson insulator in the presence of an arbitrarily small disorder\cite{lee_disordered_1985}. The clean limit superconductivity in 2D is stabilized through a Berezinskii-Kosterlitz-Thouless (BKT) transition\cite{kosterlitz_ordering_1973}. At $T=0$, simply by changing disorder, a 2D superconductor can be converted into an insulator through a quantum phase transition. Though it is not expected within the quantum theory highlighted above, a physical disorder often drives a 2D superconductor to a metallic phase, where Bosonic particles may undergo metal-like transport, before the system becomes completely insulating\cite{sacepe_quantum_2020, kapitulnik_colloquium_2019}. This happens through a gradual destruction of the superconducting critical temperature ($T_c$)\cite{garcia-barriocanal_electronically_2013, vaitiekenas_anomalous_2020, li_anomalous_2019}. However, physical disordering of a sample is irreversible and that restricts the investigation of the phase in a controlled manner. In addition to physical disordering, another way of creating a disordered superconductor can be by making nanofabricated 2D arrays of semiconducting and superconducting islands\cite{eley_approaching_2012, han_collapse_2014, bottcher_superconducting_2018} where the inter-island distance can be controlled by physically varying the distance in different samples ($ex$-$situ$). Such $ex$-$situ$ fabrication also involves variation of other key parameters, other than the inter-island distance. Here we have shown that such an array can be reproducibly constructed and controllably tuned $in$-$situ$ by an electric-field at the interfaces between the Mott insulator LaVO$_3$ and the band insulator SrTiO$_3$. Well segregated superconducting islands embedded in a semiconducting matrix are created, the island size and the inter-island distance is controlled to achieve a quantum anomalous metallic (QAM) phase simply by tuning an external electric field ($V_G$). It is known that superconductivity and ferroelectricity are intimately related in SrTiO$_3$\cite{PhysRevB.98.104505,dikin_coexistence_2011, bert_direct_2011, mohanta_phase_2014, michaeli_superconducting_2012, gastiasoro_superconductivity_2020} and such a control is achieved through the control of the ferroelectric domains on locally strained SrTiO$_3$\cite{frenkel_imaging_2017, haeni_room-temperature_2004, jyotsna_ferroelectric-like_2014, andreeva_temperature_2015, bi_electro-mechanical_2016, huang_direct_2013}. The QAM phase thus obtained displays strange disorder (or $V_G$) dependence. While the resistance of the system changes by several orders of magnitude with $V_G$, the transition temperature to the metallic phase ($T_{CM}$) shows no appreciable change over a wide electric field range.
\section*{Results and discussion}
The LaVO$_3$/SrTiO$_3$ interfaces were grown by pulsed laser deposition (PLD). The interface forms a two-dimensional electron gas (2DEG). The details of the growth and the superconducting properties of the interfaces are presented elsewhere \cite{halder_unconventional_2022}. Here we first re-visit the fragility of the superconducting phase that was discussed in \cite{halder_unconventional_2022}. It was seen that even at extremely low measurement currents (50 nA), despite the presence of the sharp superconducting transition, the low temperature phase remained resistive in absence of any electric gate ($V_G$). In the present study, as shown in Figure 1(a) and the inset, the sheet resistance at $V_G = 0$ is $\sim$ 50$\Omega$. The resistance gradually decreased in presence of a positive $V_G$ and the zero resistance state was achieved at $V_G = +$ 40 V. Based on this observation alone, the 2DEG can be modeled as a network of resistors (see Figure 1(b)) where an assembly of a number of well segregated nanoscale superconducting islands (SCI) (resistance $R_S = 0$) are connected by a transport channel (resistance $R_N \neq 0$) provided by the background\cite{wu_theory_2004}. Please see the supporting information file for the details of the models and the method to theoretically generate the $R-T$ plots (the black lines in Figure 1(a,c)). The data also indicates that when $V_G$ is gradually increased from zero, the size of the SCIs increase thereby decreasing the inter-SCI distance. This increases the fraction of $R_S$ in the network. Eventually, when the fraction is sufficiently large ($\sim$ 77\% in our network resistor model, S2 in supporting information), they come close enough to each other (within a length-scale $\xi$) giving rise to the globally phase-coherent zero resistance state.

\begin{figure}[ht]
    \includegraphics[scale = 0.175]{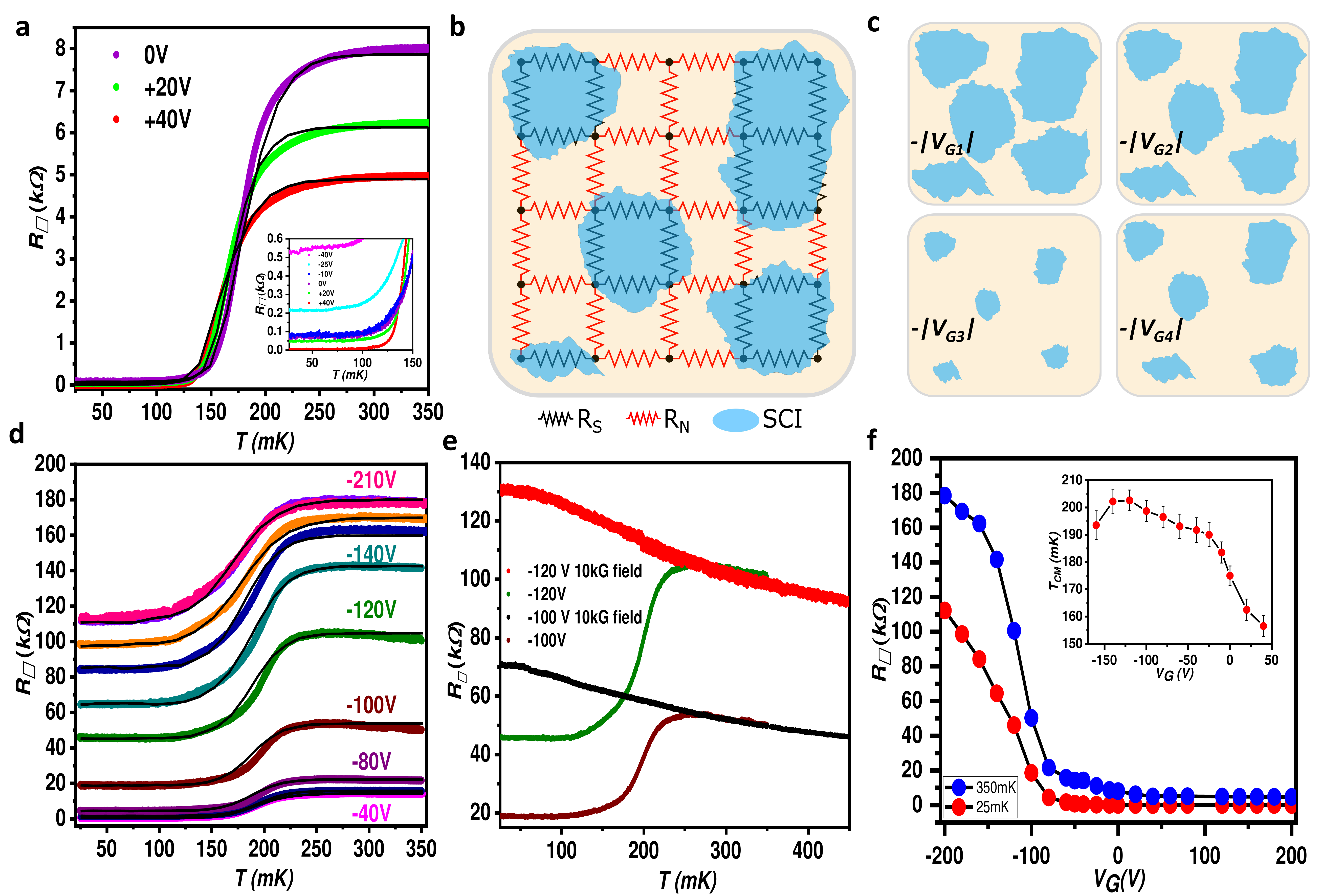}
    \caption{\textbf{Electric-field controlled phase transition.} (a) Resistance ($R$) vs temperature ($T$) plots with increasing (positive) gate voltage ($V_G$). The inset shows change in resistance below $T_{CM}$. (b) A schematic illustration of the random resistor-network model used for simulating the segregated superconducting islands at LaVO$_3$/SrTiO$_3$ interfaces. (c) A schematic demonstrating decreasing superconducting island (SCI) size and increasing inter-island distance with decreasing $V_G$. Here $\lvert{V_{G1}}\rvert$<$\lvert{V_{G2}}\rvert$<$\lvert{V_{G3}}\rvert$<$\lvert{V_{G4}}\rvert$. (d) $R$ vs $T$ plots at decreasing $V_G$ showing dramatic increase in resistance while the transition at $T_{CM}$  remains almost unaffected. Solid black lines in (a, d) are resistor-network model simulation fit. (e) $R$ vs $T$  at $V_{G}$ = -100V, -120V showing insulating behaviour of the background under applied magnetic field of $1T$. (f) $R$ vs $V_{G}$ characteristics below and above $T_{CM}$. The inset shows slight change in $T_{CM}$.}
\end{figure}

\begin{figure}[ht]
    \centering
    \includegraphics[scale = 0.175]{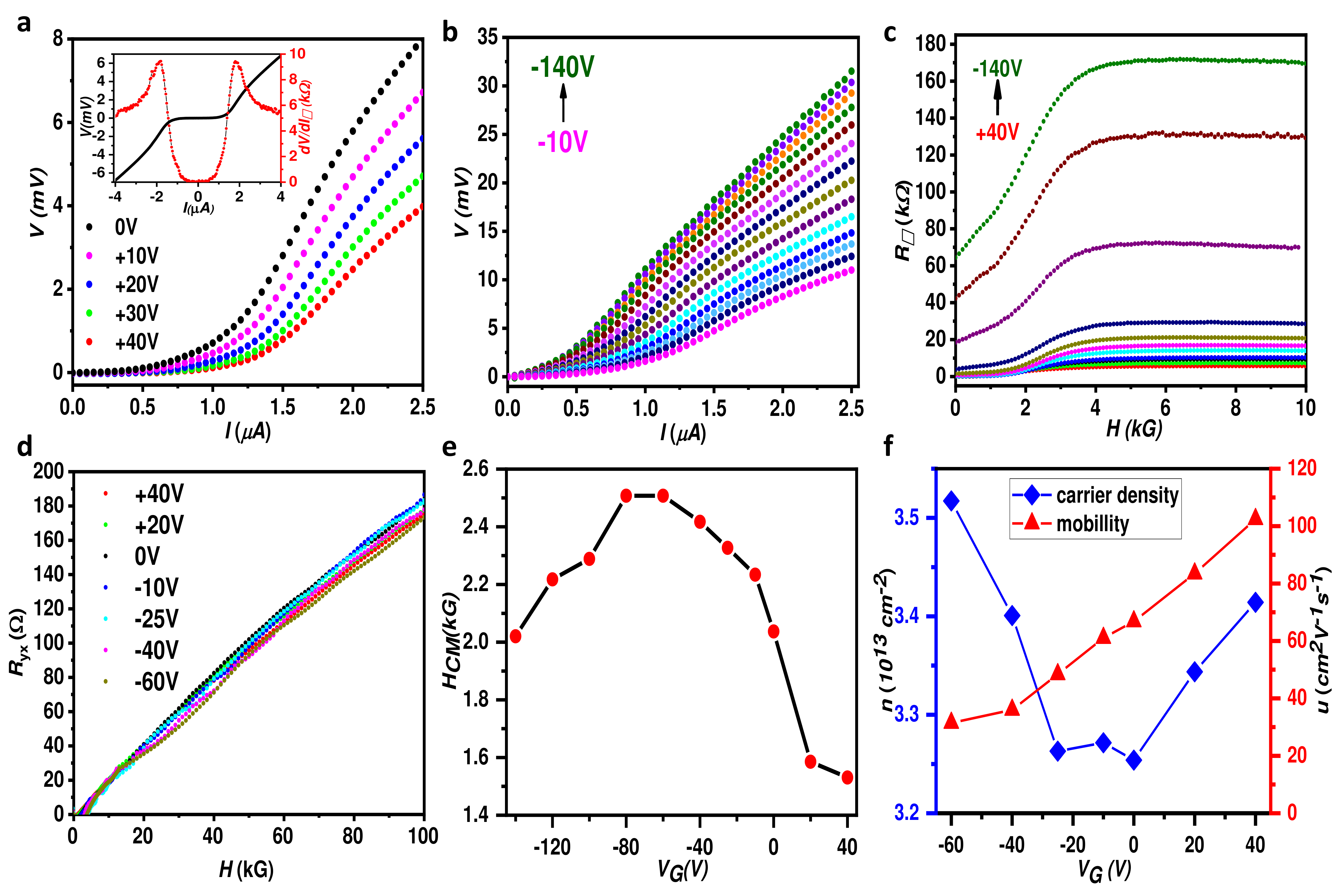}
    \caption{\textbf{Electric field dependent physical properties across the phase transition.} (a), (b) $I$ vs $V$ characteristics of the QAM phase below $T_{CM}$ with different $V_G$, positive and negative respectively.  Inset in (a) showing  $I$ vs $V$ and $dV/dI$ from $V_{G}$ = +40V. (c) Magneto-resistance $R$  vs applied perpendicular magnetic field $H$ evolution for different $V_G$. (d) Hall Measurements at different $V_G$ showing almost no change. Here $R_{yx}$ = -$R_{xy}$. (e)$H_{CM}$ values extracted from (c).  (f) Extracted carrier density and mobility values from (d).}
\end{figure}

The above observation suggests that the SCIs are electrically active and are readily modified in shape and size by electric field. This effect shows uncanny similarity with electric field control of ferroelectric domains in ferroelectrics. We had earlier shown that electric domains can be written and erased on SrTiO$_3$ and LaVO$_3$/SrTiO$_3$ controllably by an external electric field applied through a conducting cantilever in the so-called piezoresponse force microscopy (PFM) mode\cite{balal_electrical_2017}. Here it is important to note that SrTiO$_3$ is a known quantum paraelectric and exists very close to a ferroelectric phase transition\cite{haeni_room-temperature_2004}. A natural transition to a ferroelectric phase in SrTiO$_3$ is prevented by quantum fluctuations, but such a transition is easily facilitated by strain. Even at room temperature, a few atomic layers near the surface give ferroelectric response in piezo-response force microscopy (PFM)\cite{jyotsna_ferroelectric-like_2014,andreeva_temperature_2015}. It is also widely believed that the pairing glue for surprising superconductivity in dilute $\delta$-doped SrTiO$_3$ is mediated by ferroelectric fluctuations. A large volume of literature suggests that ferroelectricity and superconductivity are intimately related in SrTiO$_3$\cite{gastiasoro_superconductivity_2020, dikin_coexistence_2011, bert_direct_2011, mohanta_phase_2014, michaeli_superconducting_2012}. As the superconductivity of $\delta$-doped SrTiO$_3$ is trapped in 2D when LaAlO$_3$ or LaVO$_3$ is deposited on SrTiO$_3$, ferroelectricity also persists underneath such interfaces\cite{frenkel_imaging_2017, bi_electro-mechanical_2016, huang_direct_2013, balal_electrical_2017}. In such cases, the primary origin of interfacial strain is related to a structural phase transition of the SrTiO$_3$ substrate below 105K, from the high temperature cubic to low-temperature tetragonal phase\cite{lytle_xray_1964, gastiasoro_superconductivity_2020}. Ferroelectric domains in SrTiO$_3$ are also seen near intrinsic crystal defects like the twin boundaries, or other forms of local mechanical disorder\cite{frenkel_imaging_2017, scott_domain_2012, xu_strain-induced_2020, jang_ferroelectricity_2010}. The phase boundary separating the non-ferroelectric and the ferroelectric phases of SrTiO$_3$ is extremely narrow for expansive strain. Since the lattice constant of LaVO$_3$ (3.95 \AA) is larger than that of SrTiO$_3$, and that SrTiO$_3$ undergoes a structural phase transition below 105 K, the SrTiO$_3$ side of the LaVO$_3$/SrTiO$_3$ interfaces is significantly strained and can host strong ferroelectricity. Pronounced electric-field dependent phase switching in PFM signal was indeed seen in LaVO$_3$/SrTiO$_3$ with thin (10 monolayers) of LaVO$_3$\cite{balal_electrical_2017}. Considering the above, and noting the similarity of the observed effect with the electric field control of ferroelectric domains in ferroelectrics, it is rational to surmise that in case of LaVO$_3$/SrTiO$_3$ interfaces, the SrTiO$_3$ side of the interface hosts ferroelectric domains with out-of-plane spontaneous electric polarization. Our model would be applicable regardless of the origin of ferroelectric domains in SrTiO$_3$. When the 2DEG forms in presence of the underlying ferroelectric domains (perhaps for resolving potential polar catastrophe), puddles of high and low carrier density regions, commensurate with the local direction of electric polarization, can spontaneously form. In such a picture it is natural that the high carrier density regions undergo superconducting transition below $T_c$ and form the SCIs. With increase in $V_G$, as the ferroelectric domains increase in size, the size of the SCIs also change in a directly correlated manner and eventually they get connected to become globally superconducting (at $V_G = + 40 V$).

If the SCIs are correlated with the ferroelectric domains in SrTiO$_3$ as per the discussion above, it is naturally expected that when the direction of the electric field is reversed ($V_G$ negative), the size of the SCIs should decrease (as illustrated in Figure 1(c)) and the effective separation between the SCIs increase. That would cause a decrease in the fraction of $R_S$ leading to an overall increase in the effective network resistance. As presented in Figure 1(d), a dramatic increase in resistance, even below the superconducting transition of the individual SCIs is indeed seen when $V_G$ was gradually decreased down to -210 V. The black lines over the color data points show the simulated $R-T$ plots within the network resistor model assuming a systematic decrease in the fraction of $R_S$ in the model with decreasing $V_G$. This observation further validates the claim that the control on the size of the SCIs is obtained through the control of the ferroelectric domains in in the SrTiO$_3$ side of the LaVO$_3$/SrTiO$_3$ interfaces. Within the range of $V_G$ that we worked with, electronic doping effect was far less prominent (due to bottom gating under a 0.5 mm thick SrTiO$_3$ crystal) than the ferroelectric domain modification.


From the above discussion, regardless of the mechanism through which such control is achieved, it is clear that with $V_G$, the size of the SCIs and the distance between the SCIs change consistently. Therefore, changing $V_G$ is equivalent to changing electronic disorder in this case. This implies that with larger values of negative $V_G$, a smooth destruction of superconductivity followed by a transition to an insulating phase would be expected. Instead, we observe that for $-210 V < V_G < -40 V$, the system settles in an anomalous metallic phase where the resistance of the 2DEG drops below a temperature $T_{CM}$, as if it was tending to undergo a superconducting transition, but the system resistance saturated at low temperatures, without completing the superconducting transition. In this state, while the Bosonic Cooper pairs are phase-coherent within individual SCIs, a global phase coherence is lost. The saturation to the lowest resistance happens over a rather broad temperature range ($\sim$ 100 mK). This is attributed to a variation of superconducting transition temperature $T_c$ of the SCIs. The scheme of determining $T_c$ is given in supporting information(Figure S4). We also noted that below $V_G =$ -60 V, the resistance (below and above $T_{CM}$) increases more rapidly with further decreasing $V_G$, indicating that $V_G \sim$ -60 V corresponds to a critical point where a significant fraction of the SCIs get separated from others by a distance larger than their respective coherence length ($\xi$). From our model we found that such a separation happens when approximately 46\% of the resistors in the network become superconducting. Beyond $V_G=-$100 V, the normal state resistance (for $T>T_{CM}$) slowly increases with decreasing temperature indicating that the normal state has undergone a transition from a bad-metal to an insulator (see Figure 1(e) and S5(b)). However, the resistance below $T_{CM}$ did not show any upward curvature down to $V_G=-$ 210 V. This indicates that the system as a whole has reached a new strange quantum anomalous metallic phase below $T_{CM}$ which is seemingly similar to a so-called Bose metal\cite{phillips_elusive_2003}. For $V_G<-$100 V, when the superconductivity of all the SCIs is suppressed by a large externally applied magnetic field $H =$ 10 kG, the QAM phase gets completely destroyed and the system shows a smooth insulating behaviour down to the lowest temperature (Figure 1(e)).    

\begin{figure}[ht]
    \centering
    \includegraphics[scale = 0.5]{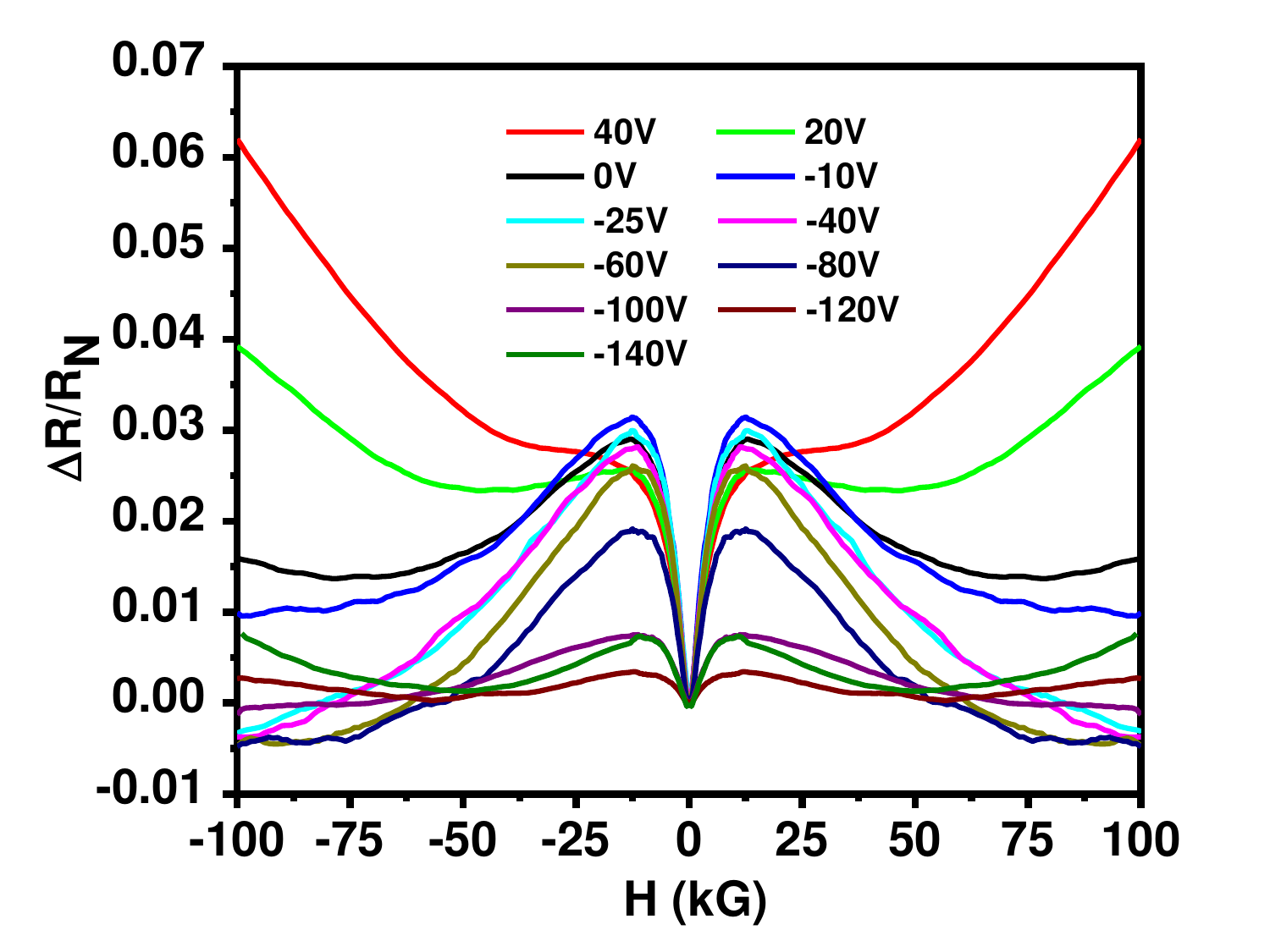}
    \caption{\textbf{Electric field dependence of longitudinal magnetoresistance (MR):} Normalized magnetoresistance measured at 350mK (above T$_{CM}$) for different $V_G$. $\Delta R = R(H)-R(0)$ and $R_N$ is normal state  resistance.}
\end{figure}

The $I-V$ characteristics of the QAM show non-linearity as the $I-V$ of the individual SCIs are non-linear due to their respective finite critical currents (Figure 2(a)). A true dissipation-less state corresponding to global superconductivity is seen for $V_G >$ +40 V (inset of Figure 2(a)). As it is seen in Figure 2(c), the QAM phase is destroyed by $H$ over a broad range of $H$. This is understood as a distribution of the superconducting upper critical field ($H_{c2}$) of the individual SCIs. While the $H$-dependent transition begins at a very low $H$ of $H_1\sim$ 100 G, the destruction of the QAM phase is seen to be complete around $H_{2}\sim$ 4 kG. $H_1$ gives the lower limit of the $H_{c2}$ for an SCI. $H_2$ corresponds to the highest $H_{c2}$ among all the SCI. This gives a coherence length $\xi\sim$ 30 nm. The average critical field for this transition ($H_{CM}$) varied between 1.4 kG and 2.4 kG (Figure 2(e) leading to a variation in the average coherence length ($\xi_{av}$) between 40 nm and 55 nm. Since $H_{CM}$ is $V_G$-dependent, $\xi_{av}$ is also $V_G$-dependent.   

Although the resistance values calculated from our network-resistor model are consistent with experimental observations, they do not fully explain the Figure 1(f) inset. The dome-shaped curve for T$_c$ vs V$_G$ is obtained, even though the change in T$_c$ is very small. The interesting observation here is the reversal of the under-doped and over-doped regimes. The under-doped regime is accessed at positive V$_G$, whereas the over-doped regime is found at negative V$_G$. This phenomenon suggests that there is carrier redistribution between the superconducting and non-superconducting regions during the process. As the island sizes increase, the carrier density of the islands gets diluted, and a reverse effect occurs when the islands shrink. To understand the details of the mechanism, further theoretical and experimental investigations are required. However, we believe that a similar effect may be observed in other SrTiO$_3$-based interfaces where there is a similar lattice mismatch (i.e., experiencing expansive strain) at the interfaces. Remarkably, as seen in Figure 1(f), while the resistance of the 2DEG below and above $T_{CM}$ changed by almost two orders of magnitude within a large range of $V_G$, $T_{CM}$ changed by only 10\% (see inset of Figure 1(f)), indicating that the distribution of superconducting $T_c$ over the SCIs do not vary appreciably with $V_G$. This implies that with changing $V_G$ the carrier density does not change significantly within the SCIs. In order to confirm this we have performed $V_G$ dependent Hall measurements above $T_{CM}$. As shown in Figure 2(d), the magnetic field dependence of the Hall signal is nonlinear indicating multiple bands participating in the transport\cite{hotta_polar_2007}. This is consistent with the band structure calculations earlier reported by us, where it was seen that multiple electronic bands take part in transport at the LaVO$_3$/SrTiO$_3$ interfaces. An analysis of the Hall data (Figure 2(f)) reveals that the carrier density of the whole system changes only by 5\% for -60<$V_G$<+40. The change in resistance with $V_G$ is predominantly governed by comparatively larger change in mobility. It should be noted that the estimate of mobility is not exact as the magnetic field dependence of the Hall resistance is non-linear due to multiple bands participating in transport\cite{halder_unconventional_2022}. As shown in Figure 3, the longitudinal magnetoresistance (MR) of the normal state (above $T_{CM}$) also varied significantly with $V_G$. There is a generic difference in the qualitative shape of the MR for positive and negative $V_G$ respectively. The shape of the MR vs. $H$ plot changes when the sign of $V_G$ is reversed. At $V_G = 40 V$, for which we have obtained the zero resistance, the MR is always positive but forms a shoulder at $\pm$ 2.5 kG. Such overall anomalous variation of MR (transverse or longitudinal) for different $V_G$ is often seen to originate from a Berry curvature effect in material systems with strong spin-orbit coupling (SOC)\cite{kumar_observation_2021, caviglia_tunable_2010, ma_tunable_2014} and is surprisingly similar to the chiral anomaly effects in transverse magnetoresistance of Weyl semimetals\cite{wang_gate-tunable_2016}. Since the SOC is expected to be strong in LaVO$_3$/SrTiO$_3$, the non-monotonic MR in Figure 3 may be a consequence of that. The exact origin of such a non-trivial MR and the role of SOC in the formation of the SCIs and their $V_G$ dependence need further theoretical investigation which is beyond the scope of this article. 
\section*{Conclusion}
In conclusion, we have achieved unprecedented control over the distribution of nanometer scale superconducting islands in an oxide 2DEG simply by tuning the electric field dependent ferroelectric domain size. This is a demonstration of a reversible way of controlling inhomogeneity $in$-$situ$ in two dimensions and stands out in the research of inhomogeneous materials where, so far, changing inhomeogeneity essentially meant changing the material itself. With such control, we have realized an electric field induced quantum phase transition from a 2D superconductor to an anomalous strange metallic phase. In this phase, it appears that phase-incoherent Bosons possibly take part in metal-like transport as in a Bose metal, without undergoing a global condensation. This provides an important insight into the mechanism of emergence of superconductivity in SrTiO$_3$ based low-carrier density superconducting systems. The data suggests that in such systems there are multiple energy scales where phase-incoherent Bosonic pairs form before global phase coherence to a superconducting condensate happens. Controlling nanoscale superconducting islands by tuning ferroelectricity with an external electric field provides a new avenue of controlling electronic disorder in low dimensions and this will spur research to realize new quantum phases in low-dimensional quantum materials through $in$-$situ$ ferroelectric domain engineering.

\section*{Methods}

The LaVO$_3$/SrTiO$_3$ oxide hetrostructure was grown by pulsed laser deposition. Growth was monitored by $in$-$situ$ RHEED (reflection high-energy electron diffraction). Initial characterization measurements of the 2DEG formed at the interface were done in a Quantum Design PPMS (Physical Property Measurement system) as mentioned in \cite{halder_unconventional_2022}.

All the measurements reported here were performed in a dilution refrigerator (custom designed, built by Janis ULT) with sample stage (home-built, made of OFHC (oxygen free high-thermal conductivity) copper) temperature going to 28 mK. The cryostat is equipped with a 160 kG bottom loaded superconducting magnet. The homogeneity of the applied magnetic field to the sample stage is $\pm$0.1\% over 10 mm DSV. The sample was mounted on custom designed Rogers RO4003C PCB (on sample stage with silver epoxy and electrically connected  in Hall geometry to the gold coated PCB tracks using ultrasonic wire (1 mil Al) bonded contacts. The whole stage was mounted inside a custom designed gold coated OFHC copper sample holder inside a silver cradle in  direct contact with the mixing chamber using a vacuum manipulator. 

The AC resistance measurements were performed using a SR830 lock-in amplifier in conventional four probe method. An excitation current of 50 nA at 16.356 Hz was given. For AC Hall measurements at 350 mK, excitation current of 1.5$\mu$A was given. The time constant of the lock-in amplifier was set to 300 ms. Measurements were also repeated using 10kHz low pass filter, no change was found.

For varying the sample temperature, a home built heater using 36AWG Phosphor bronze wire was mounted on the sample stage and was controlled by Keithley 6221 current source. The heater current was increased in the steps of 2$\mu$A with wait time of 17 seconds at each step. A Ruthenium Oxide (RuOx) thermometer read by Lakeshore 372 controller was mounted on the sample stage.

For DC $V-I$ measurements, Keithley 6221 current source and Keithley 2000 multi-meter were used. The DC current was increased in the steps of 0.04 $\mu$A  and wait time at each step was 3 seconds.
For applying gate voltages, Keithley 2400 Source-meter was used. DC voltage was increased in step size of 50mV and wait time at each step was 45 seconds.

\section*{Supporting Information}
Random resistor network model, Determination of T$_c$ and error estimation, Additional data.
\subsection*{Data Availability}
Data are available from the corresponding author upon reasonable request.

\section*{Acknowledgements}
We thank Dr. Tanmoy Das for useful discussions. MG thanks the Council of Scientific and Industrial Research (CSIR), Government of India, for financial support through a research fellowship (Award No. \textbf{09/947(0227)/2019-EMR-I}). NSM thanks UGC for Senior Research Fellowship (SRF). GS acknowledges financial assistance from the Department of Science and Technology (DST), Govt. of India through Swarnajayanti fellowship (grant number: \textbf{DST/SJF/PSA-01/2015-16}) and the Science and Engineering Research Board (SERB), Govt. of India (grant number: \textbf{CRG/2021/006395}).

\subsection*{Author contributions} 
SH, MG and NSM carried out the ultra-low temperature experiments with help from GS. SH and MG fabricated the ultra low-temperature measurement hardware. AK and SC fabricated the LaVO$_3$/SrTiO$_3$ heterostructures and characterized them. SG and SK performed the model simulations. GS wrote the manuscript with help from all the co-authors.

\section*{Competing interests}
The authors declare no competing financial interests.






\clearpage
\newpage
\onecolumngrid
\begin{center}
	\textbf{\large Supporting Information: Electrically controlled quantum transition to an anomalous metal in 2D}

Soumyadip$^a$ Halder, Mona$^a$ Garg, Shreekant Gawande, Nikhlesh Singh Mehta, Anamika Kumari, Suvankar Chakraverty, Sanjeev Kumar, Goutam Sheet$^*$

Department of Physical Sciences, Indian Institute of Science Education and Research (IISER) Mohali, Sector 81, S. A. S. Nagar, Manauli, PO 140306, India

Quantum Materials and Devices Unit, Institute of Nano Science and Technology,
Sector-81, Punjab, 140306, India.
\end{center}
\def\thefootnote{$^*$}\footnotetext{goutam@iisermohali.ac.in}
\def\thefootnote{$^a$}\footnotetext{These authors contributed equally to the work.}

\setcounter{equation}{0}
\setcounter{figure}{0}
\setcounter{table}{0}
\setcounter{page}{1}
\makeatletter
\renewcommand{\theequation}{S\arabic{equation}}
\renewcommand{\thefigure}{S\arabic{figure}}
\renewcommand{\bibnumfmt}[1]{[S#1]}
\renewcommand{\citenumfont}[1]{S#1}

\maketitle

\section*{Random resistor network model:}
\begin{figure}[ht]
    \centering
     \includegraphics[scale=0.4]{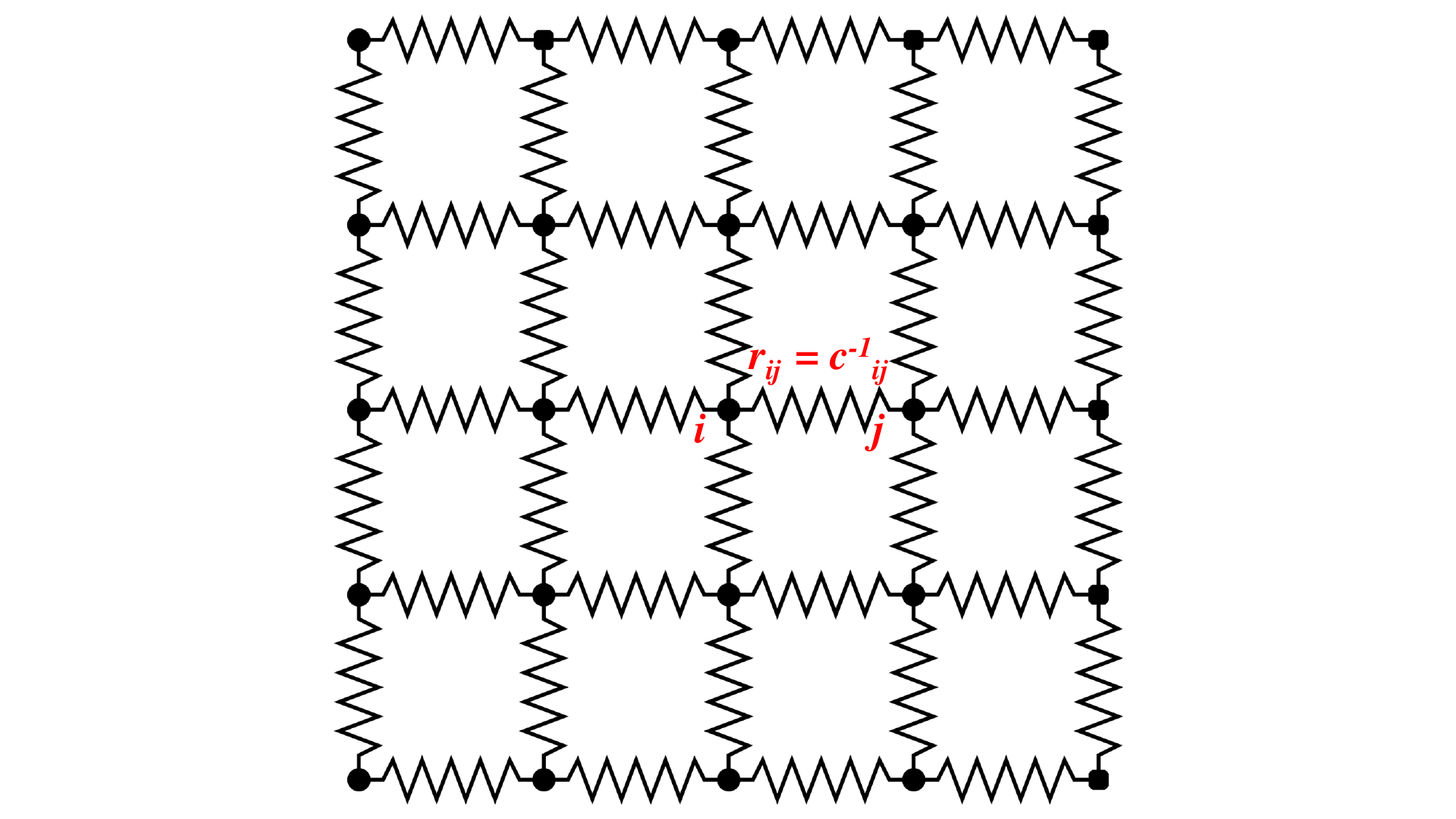}
     \caption{Schematic of a resistor network}
\end{figure}

We model the results on a  30$\times$30 square lattice of nearest neighbour resistor links with periodic boundary conditions. Effective resistance between any two nodes $\alpha$ and $\beta$ on the lattice is given by \cite{wu_theory_2004}
\begin{eqnarray}
    R_{\alpha \beta} = \sum \frac{1}{l_i} |w_{i\alpha} - w_{i\beta}|^2
\end{eqnarray}
where $l_i$ are eigenvalues of Laplacian matrix L and $\omega_{ij}$ are components of eigenvector $W_i$ of the matrix L. The matrix L is obtained from Kirchhoff law written in the matrix form
\begin{eqnarray}
    \sum_{i \neq j}^{N} c_{ij} (V_i - V_j) = I_i \nonumber \\
    LV = I
\end{eqnarray}
where $c_{ij} = \frac{1}{r_{ij}}$ and $r_{ij}$ is resistance of the link between two nodes $i$ and $j$ in link. We apply eq.(1) to get resistance between two nodes on opposite corners on the diagonal of lattice. As discussed in experimental data analysis, we use two resistance values to model superconducting and normal states. Normal state resistance is taken to be constant against the temperature change as its variation is very small compared to superconducting transition. Constant resistance will suffice in explaining qualitative behaviour of normal state resistance. We consider a function that mimic the smooth superconducting transition against the temperature for which input variable is set such that it gives transition at $T_c$ found in experiments. We set $p$ to be fraction of the links that are superconducting, remaining links being of normal state. Then we set the resistance above $T_c$ by varying the resistance of normal state links. Once we have fit the normal resistance, we vary the fraction of superconducting links, to arrive at low temperature resistance in below $T_c$. We find that for higher positive $V_g$, $p$ turns out to be large, conversely for high negative $V_g$, $p$ takes low value. These results support experimental data.
Following is the plot of SC fraction vs normal state resistance values obtained from numerical solutions on 2D network-resistor model in agreement with the experimental values. 
\begin{figure}[ht]
    \centering
     \includegraphics[scale=0.5]{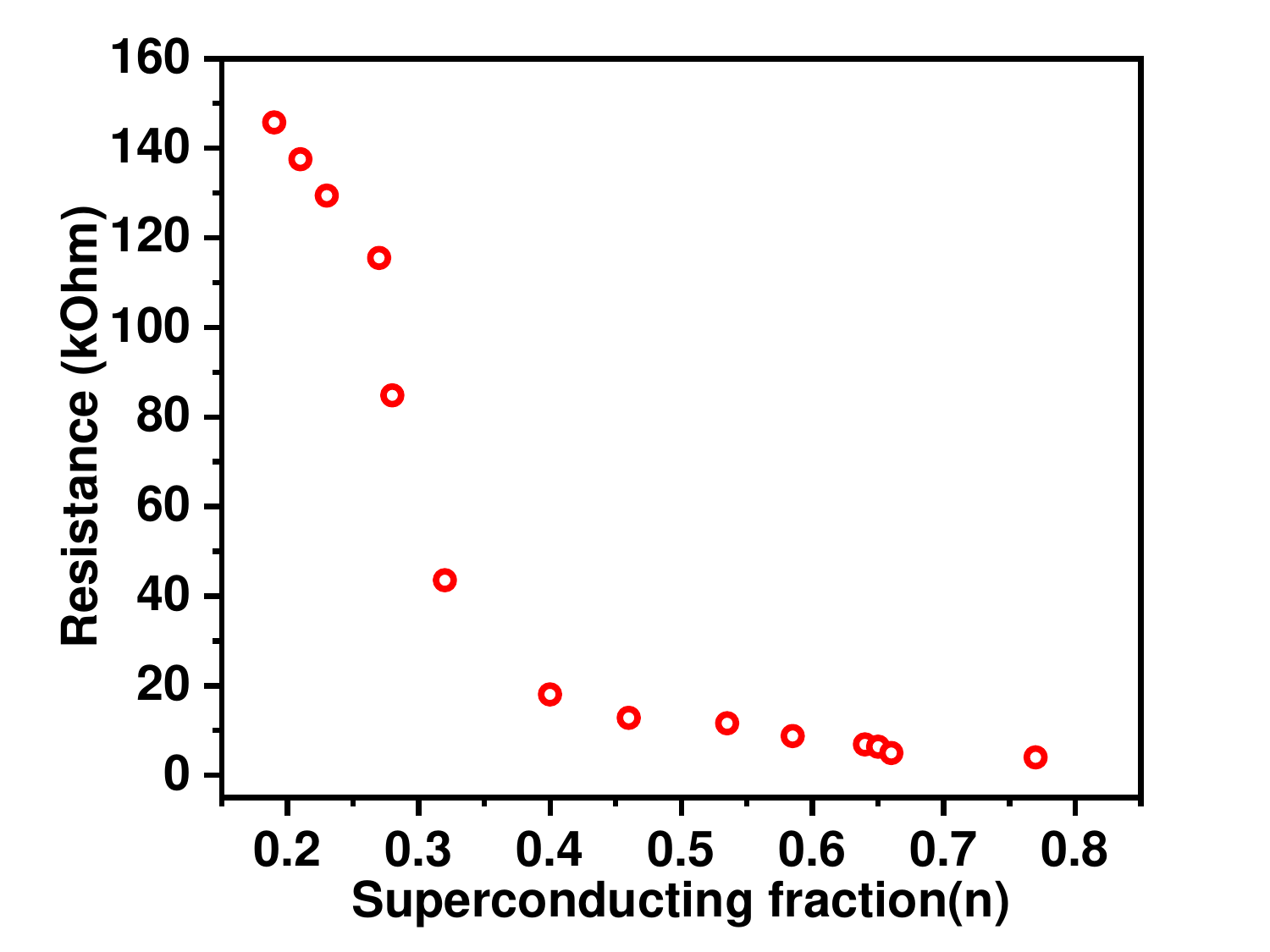}
    \caption{Fitting parameters of Network-resistor model simulation in agreement with experimental results.}
\end{figure}

It should be noted that the purpose of this data is to obtain the value of the SC fraction that generates the normal state resistance in agreement with the experimental values. This graph should not be read as the net resistance obtained for a given value of SC fraction. Therefore, this is not the graph to check the percolation threshold in a 2D random network. Instead, the above plot can be used to infer the correlations in the locations of the building blocks. If the SC links show a tendency to cluster together, then the percolation threshold will shift to values larger than 0.5. This is indeed what we find. We have explicitly verified that our calculations are consistent with the expected percolation threshold of 0.5 if we assume uncorrelated random distributions of resistors.
\begin{figure}[ht]
    \centering
     \includegraphics[scale=0.5]{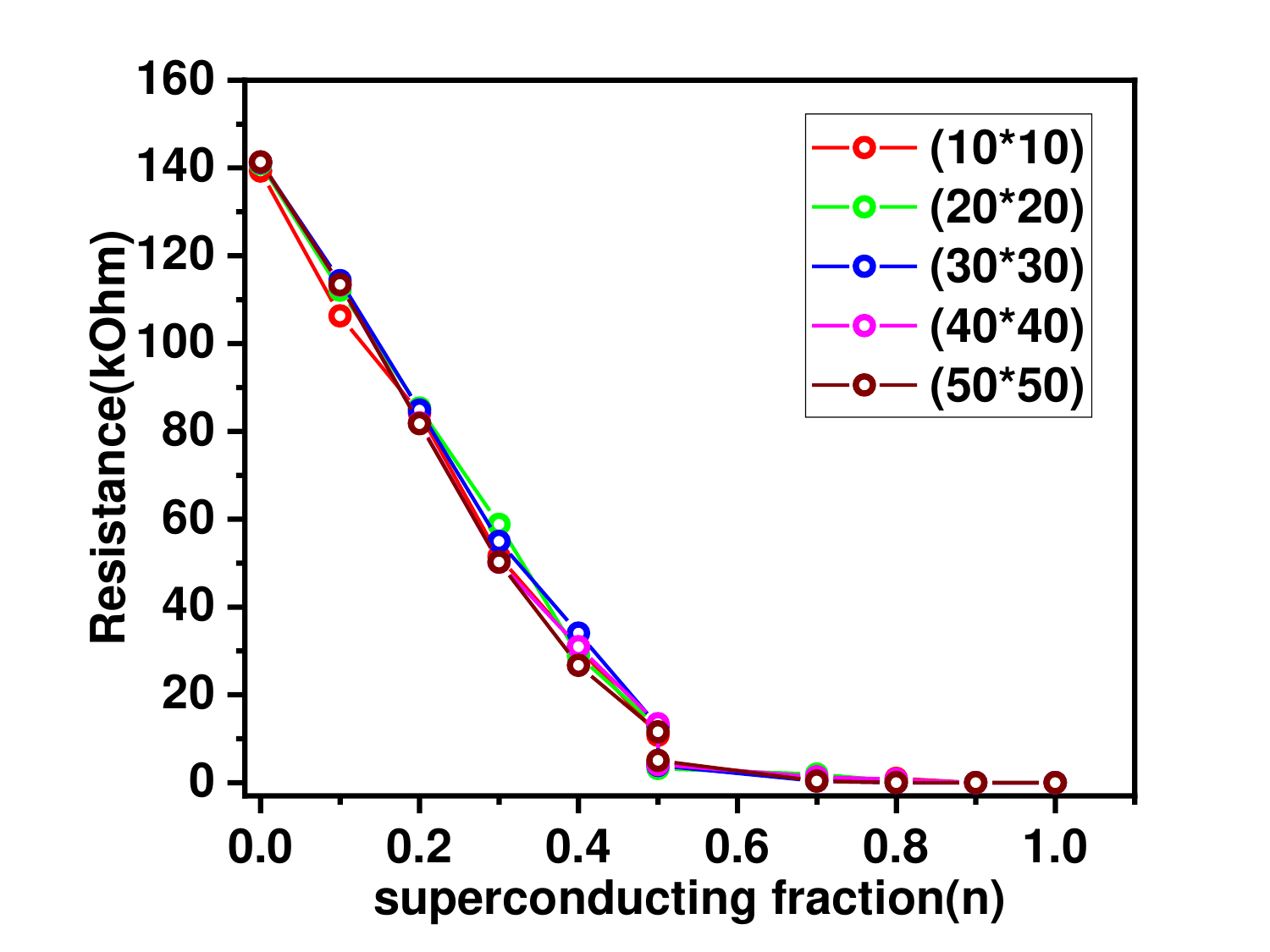}
    \caption{Calculated resistance values for various sizes for varying superconducting fraction using uncorrelated random resistor network model.}
\end{figure}

\section*{Determination of T$_c$ and error estimation:}
\begin{figure}[ht]
    \centering
     \includegraphics[scale=0.5]{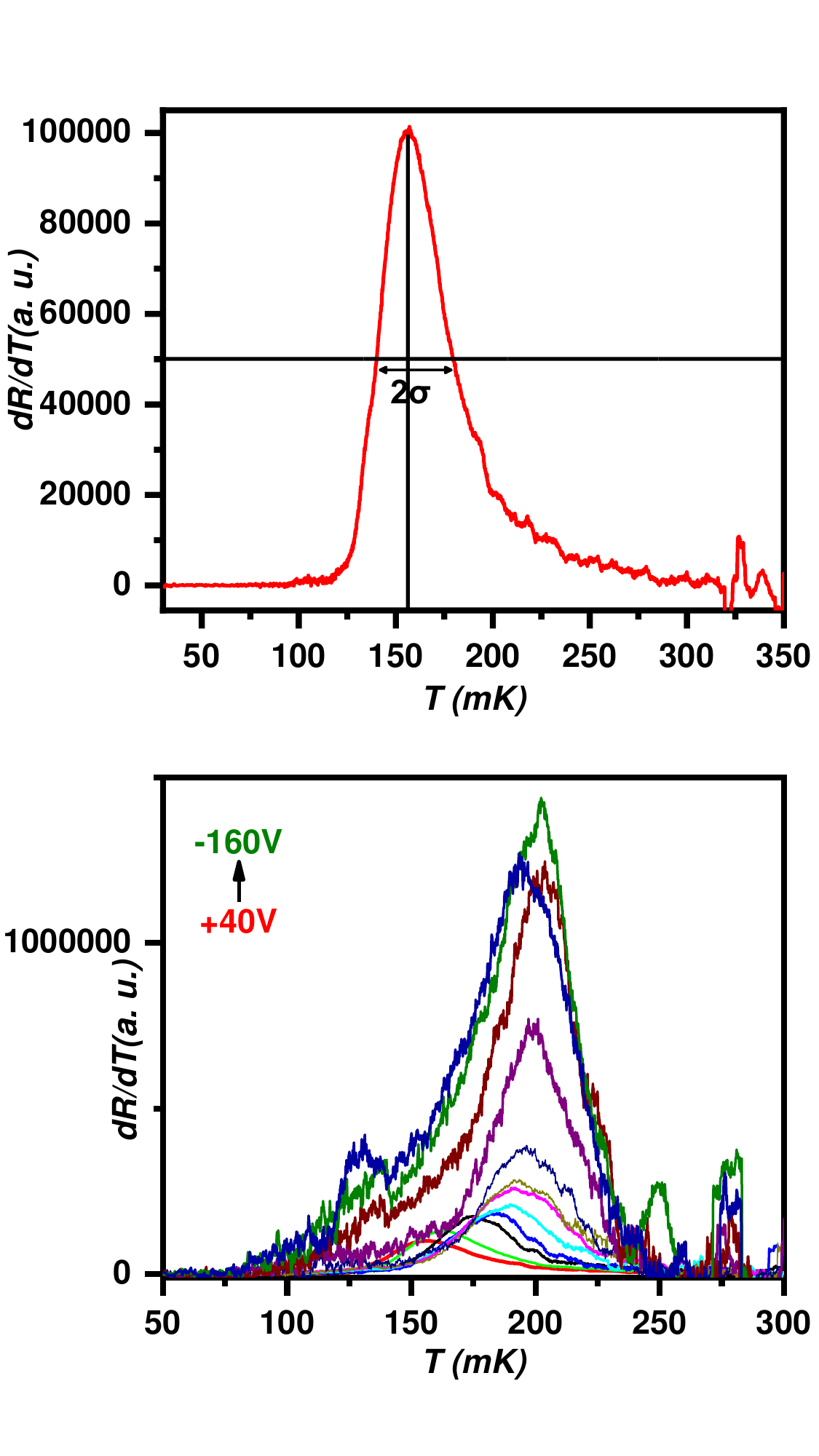}
    \caption{(a, b) $dR/dT$ vs $T$ derived from the $R$-$T$ data taken at V$_G$ = +40V and at different gate voltages respectively at 0kG.}
\end{figure}
The $R$-$T$ data was smoothed with 50 point adjacent averaging and differentiated to plot w.r.t. $T$ (as shown in Figure 1) for each $R$-$T$ taken at gate voltages. The peak value was defined as superconducting transition temperature. The error ($err$) in $T_c$ is calculated from the spread in the peak along the temperature axis. The spread ($\sigma$) of the transition is defined as half of full-width of peak at half maxima. $\pm20\%$ of $\sigma$ for each $R$-$T$ data is taken as the error in $T_c$.
\pagebreak
\section*{Additional data:}
\begin{figure}[ht]
    \centering
     \includegraphics[scale=0.6]{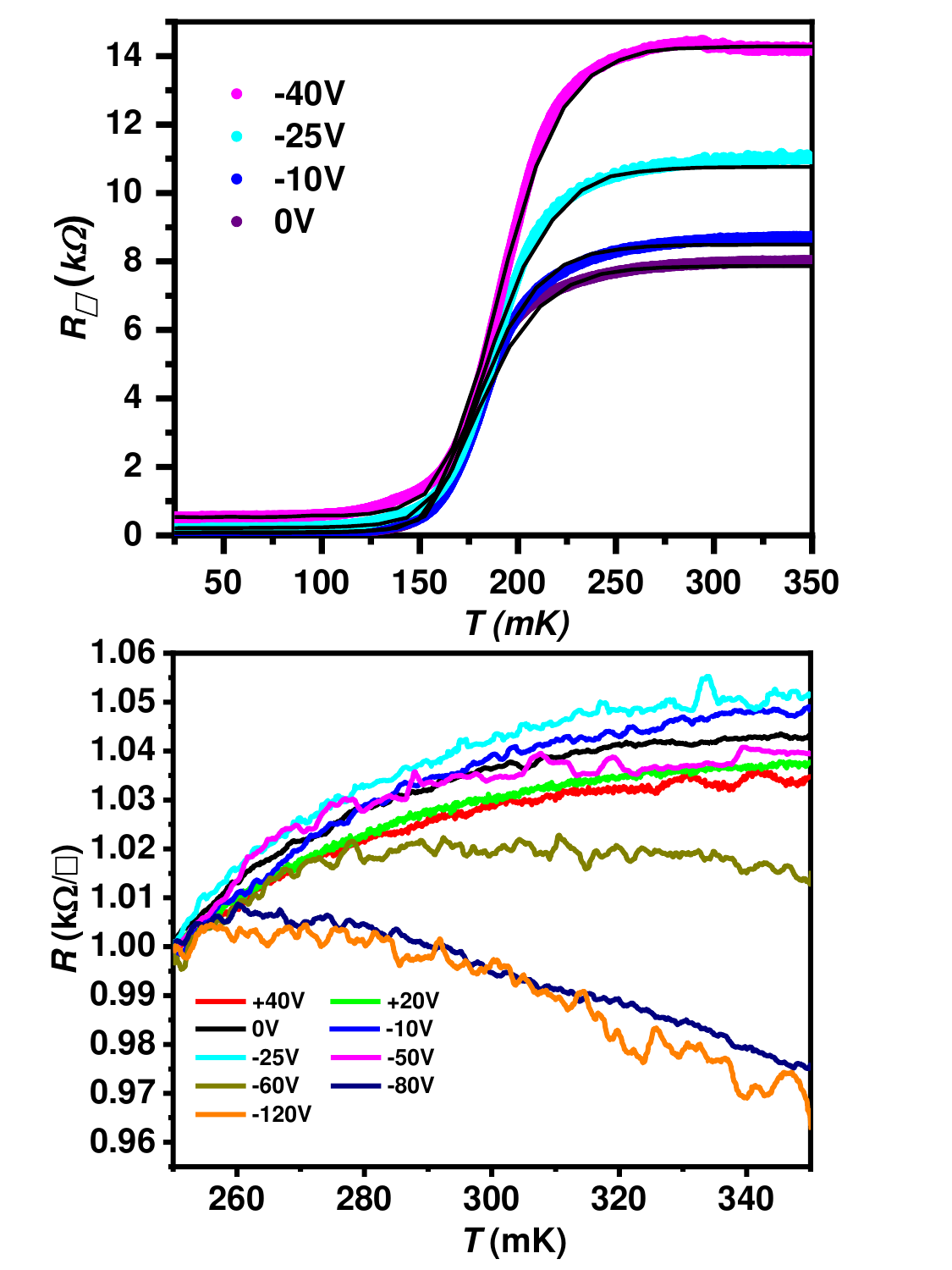}
    \caption{(a) Resistance ($R$) vs temperature ($T$) plots with decreasing (negative) gate voltage ($V_G$). (b) Resistance ($R$) vs temperature ($T$)  characteristics above $T_c$ showing insulating behaviour of normal state.}
\end{figure}
\pagebreak

\end{document}